\documentstyle[11pt]{article}
\begin{document}
\renewcommand{\refname}{References.}
\newcommand{\Nf}{N_{\!f}}
\newcommand{\half}{\mbox{\small{$\frac{1}{2}$}}}
\newcommand{\partialslash}{\partial \!\!\! /}
\newcommand{\kslash}{k \!\!\! /}
\title{Anomalous dimension of non-singlet Wilson operators at $O(1/N_{\!f})$ in
deep inelastic scattering.}
\author{J.A. Gracey, \\ Department of Applied Mathematics and Theoretical
Physics, \\ University of Liverpool, \\ P.O. Box 147, \\ Liverpool, \\ L69 3BX,
\\ United Kingdom.}
\date{}
\maketitle
\vspace{5cm}
\noindent
{\bf Abstract.} We use the large $\Nf$ self consistency formalism to compute
the $O(1/\Nf)$ critical exponent corresponding to the renormalization of the
flavour non-singlet twist two Wilson operators which arise in the operator
product expansion of currents in deep inelastic processes. Expanding the
$d$-dimensional expression in powers of $\epsilon$ $=$ $(4-d)/2$ the
coefficients of $\epsilon$ agree with the known two loop structure of the
corresponding renormalization group function and we deduce analytic expressions
for all moments, $n$, at three and higher orders in perturbation theory in the
$\overline{\mbox{MS}}$ scheme at $O(1/\Nf)$.

{\vspace{-18cm}
\hspace{10cm}
{\bf {LTH-322}}}
\newpage
The quantum field theory describing the dynamics of strongly interacting
particles is quantum chromodynamics, (QCD), which is an asymptotically free
gauge theory, \cite{1,2}. In other words at increasingly higher energies the
theory behaves more and more like a free theory and therefore the machinery of
perturbative quantum field theory can be applied to gain an accurate
understanding of physical processes. For example, the phenomenology of deep
inelastic scattering can be explored and the one and two loop QCD corrections
to parton model predictions can be computed which are found to be in good
agreement with current experiments. (See, for example, \cite{3}.) With the
increase in energy that will soon become available at LHC there would now
appear to be a specific need to begin to examine subsequent three loop
corrections to these processes, since the error bars on experimental results
will be refined substantially.

One of the central ingredients of the scattering formalism in gauge theories
is the operator product expansion, (OPE), of electromagnetic and other
currents, \cite{4}, which are composite operators of the fields of the standard
model and QCD lagrangians. The decomposition of the Wilson OPE is in terms of
an infinite sum of composite operators which are either flavour non-singlet or
singlet, \cite{4}. However, these operators require renormalization. In other
words, radiative corrections generate a non-zero anomalous dimension,
$\gamma^{(n)}_{\mbox{\footnotesize{NS}}}(g)$, for the $n$th moment of, say,
each non-singlet operator, where $g$ is the coupling constant which will be
defined later. Knowledge of $\gamma^{(n)}_{\mbox{\footnotesize{NS}}}(g)$ and
the anomalous dimensions of the singlet operators are essential for
determining, for example, corrections to sum rules. Although the Wilson
coefficients are also required for this, where these are the coefficients which
appear with each operator of the OPE, they depend on the specific process
involved, unlike the anomalous dimensions, and therefore require separate
treatment, \cite{3}. Like the functions of the renormalization group equation,
(RGE), $\gamma^{(n)}_{\mbox{\footnotesize{NS}}}(g)$ are calculated order by
order in perturbation theory. Specifically one renormalizes the operator as a
zero momentum insertion in some Green's function. As a result the one loop
structure of non-singlet operators,
$\gamma^{(n)}_{\mbox{\footnotesize{NS}}}(g)$, was given in \cite{4} whilst the
two loop analysis was carried out in [5-7]. (Two loop results for the singlet
functions have been deduced in [4,8-11].) So it is the perturbative structure
of these functions which must be determined at three loops in order to refine
the theoretical understanding of physical processes.

Such calculations, however, are formidable in their computational complexity
due to the huge number of Feynman diagrams to be analysed. Indeed the two loop
results were performed using computer algebra packages. Therefore it is also
important to have an independent way of calculating information on the
perturbation series which agrees with known calculations and, equally, provides
additional non-trivial information at much higher orders. In this letter we
introduce such a technique for computing the anomalous dimensions of physical
operators involved in the OPE of deep inelastic scattering, though for
simplicity we will only concentrate on the non-singlet case to illustrate the
power and beauty of the method. (As the Wilson coefficients are process
dependent they are beyond the scope of this letter.) The formalism is based on
the large $\Nf$ self consistency programme which was first introduced in
\cite{12,13} for the $O(N)$ bosonic $\sigma$ model. To deduce information on
the perturbative structure of any function of the RGE, one computes the
relevant critical exponent at the non-trivial zero, $g_c$, of the
$d$-dimensional $\beta$-function of QCD. There the theory is finite and the
propagators of the fields and Green's functions have a simple power law or
conformal structure. The associated critical exponent is, by universality, a
function of the spacetime dimension, $d$, and any internal parameters. For the
present letter, these will be $\Nf$, the number of quark flavours, and $N_c$
the number of colours. The benefit of proceeding in this fashion is that the
RGE simplifies because of $\beta(g_c)$ $=$ $0$, $g_c$ $\neq$ $0$, to the extent
that the $d$-dimensional critical exponent is simply related to the
corresponding RGE function evaluated at the critical coupling. (See, for
example, \cite{14}.) Therefore, if one computes in some approximation, which
will be large $\Nf$ here, then knowledge of the critical exponent in
$d$-dimensions means calculating order by order in $1/\Nf$ one can deduce the
coefficients of the perturbative function away from criticality. At low orders
of $g$ the coefficients deduced from the exponent will agree with those
calculated in the mass independent $\overline{\mbox{MS}}$ scheme which uses
dimensional regularization. Clearly this will be a powerful way of proceeding
since one will always be able to obtain information to all orders in $g$ at
each level in $1/\Nf$. Indeed it is one aim of this letter to draw attention to
this alternative method of calculating the perturbation series of useful
physical quantities.

We recall for the interested reader that the earlier applications of the large
$\Nf$ exponent programme to four dimensional gauge theories included the
evaluation of the electron anomalous dimension at $O(1/\Nf^2)$, \cite{15}, the
QED $\beta$-function at $O(1/\Nf)$, \cite{16}, and the electron mass anomalous
dimension at $O(1/\Nf^2)$, \cite{17}. As the latter involved the computation of
the exponent of the mass operator $\bar{\psi}\psi$ which is also composite, we
have used \cite{17} as a foundation for the present work. More recently, the
quark, gluon and ghost anomalous dimensions have been deduced at $O(1/\Nf)$ in
\cite{18} which are in agreement with the previous three loop results of
[1,19-22].

To proceed with the critical point approach we recall the form of the QCD
lagrangian we use is
\begin{eqnarray}
L &=& - \, \frac{(F^a_{\mu\nu})^2}{4e^2} + i \bar{q}^{iI}\partialslash
q^{iI} + A^a_\mu\bar{q}^{iI}T^a_{IJ}\gamma^\mu q^{iJ}
- \frac{1}{e^2} f^{abc} \partial_\mu A^a_\nu A^{\mu b} A^{\nu c}
\nonumber \\
&&-~ \frac{1}{4e^2} f^{abc} f^{ade} A^b_\mu A^c_\nu A^{\mu d} A^{\nu e}
- \frac{1}{2\xi e^2} (\partial_\mu A^{\mu a})^2 \nonumber \\
&&-~ \partial^\mu \bar{c}^a \partial_\mu c^a + f^{abc} \partial_\mu \bar{c}^a
c^b A^{\mu c}
\end{eqnarray}
where $q^{iI}$ is the quark field, $1$ $\leq$ $i$ $\leq$ $\Nf$, $1$ $\leq$
$I$, $J$ $\leq$ $N_c$, $A^a_\mu$ is the gluon field with $1$ $\leq$ $a$ $\leq$
$N^2_c$ $-$ $1$, $F^a_{\mu\nu}$ $=$ $\partial_\mu A^a_\nu$ $-$
$\partial_\nu A^a_\mu$, $c^a$ and $\bar{c}^a$ are the ghost fields, $\xi$ is
the covariant gauge parameter, $T^a_{IJ}$ are the generators of $SU(N_c)$ which
has structure constants $f^{abc}$ and $e$ is the coupling constant. The
location of $g_c$ in the neighbourhood of which we will base our analysis is
given by $\beta(g_c)$ $=$ $0$, where to one loop in $d$-dimensions, \cite{1},
\begin{equation}
\beta(g) ~=~ (d-4)g + \left[ \frac{2}{3} T(R)\Nf - \frac{11}{6}C_2(G)
\right]g^2 + O(g^3)
\end{equation}
where the dimensionless coupling is $g$ $=$ $(e/2\pi)^2$, in the notation
of \cite{23}, and
\begin{equation}
\mbox{Tr}(T^aT^b) ~=~ T(R)\delta^{ab} ~~,~~ T^aT^a ~=~ C_2(R)I ~~,~~
f^{acd}f^{bcd} ~=~ C_2(G) \delta^{ab}
\end{equation}
are the Casimirs of a general classical Lie group. For $SU(N_c)$, $T(R)$
$=$ $\half$, $C_2(R)$ $=$ $(N^2_c-1)/2N_c$ and $C_2(G)$ $=$ $N_c$. Therefore
from (2)
\begin{equation}
g_c ~=~ \frac{3\epsilon}{T(R)\Nf} + O \left( \frac{1}{\Nf^2}\right)
\end{equation}
where $d$ $=$ $4$ $-$ $2\epsilon$ but it turns out that to deduce the
coefficients for $\gamma^{(n)}_{\mbox{\footnotesize{NS}}}(g)$ from the exponent
$\eta^{(n)}$ $\equiv$ $\gamma^{(n)}_{\mbox{\footnotesize{NS}}}(g_c)$ we
calculate at $O(1/\Nf)$ here, the $O(1/\Nf^2)$ corrections of (4) are not
required.

The Wilson operators whose critical exponent $\eta^{(n)}$ we will compute are
the twist $2$ flavour non-singlet operators, \cite{4},
\begin{equation}
{\cal O}^{\mu_1 \ldots \mu_n, \pm}_{\mbox{\footnotesize{NS}},a} ~=~
\half i^{n-1} {\cal S} \bar{q}^I \gamma^{\mu_1} D^{\mu_2} \ldots D^{\mu_n}
T^a_{IJ} (1\pm\gamma^5) q^J - \mbox{trace terms}
\end{equation}
where $D^\mu$ $=$ $\partial^\mu$ $+$ $iT^a A^{\mu a}$ and ${\cal S}$ denotes
the symmetrization of the Lorentz indices. At leading order in large $\Nf$ it
will turn out that the $\gamma^5$ term will be completely passive and the same
result obtained for either sign. So we will suppress the index $\pm$ for the
moment. To determine the anomalous dimensions one first of all removes the
symmetrization and trace terms by introducing a null vector $\Delta^\mu$ with
$\Delta^2$ $=$ $0$ and multiplies
${\cal O}^{\mu_1\ldots\mu_n}_{\mbox{\footnotesize{NS}}}$ by $\Delta_{\mu_1}
\ldots \Delta_{\mu_n}$ to simplify the number of insertions to be made in the
Green's function, $\langle q {\cal O}_{\mbox{\footnotesize{NS}}} \bar{q}
\rangle$, \cite{4}. Next as the anomalous dimension
$\gamma^{(n)}_{\mbox{\footnotesize{NS}}}(g)$ is gauge independent one chooses
to compute in the Feynman gauge to reduce the algebra involved in the
calculation. As a consequence of both these steps the insertion of the operator
(5) requires the development of new Feynman rules.  These have been derived in
\cite{4} for the one loop case and in \cite{5} for the $2$-loop calculations.
We will use the former rules for the large $\Nf$ computation since at leading
order the only graphs which contribute are those of fig. 1, where the dotted
line corresponds to quark fields. The circle with cross denotes the zero
momentum insertion of ${\cal O}_{\mbox{\footnotesize{NS}}}$. (In the singlet
calculation there would be additionally two $2$-loop graphs with a closed quark
loop, \cite{8}, contributing at this order in $1/\Nf$, but these are zero in
the non-singlet case.)

In the perturbative calculation one computes each graph of fig. 1 with the
conventional quark and Feynman gauge gluon propagators, $\kslash/k^2$ and
$\eta_{\mu \nu}/k^2$ respectively, in dimensional regularization and absorbs
the simple poles in $\epsilon$ minimally into the appropriate renormalization
constant. In the critical point approach one uses an alternative strategy.
Near $g_c$, (4), the propagators of the fields obey asymptotic scaling and
therefore satisfy a simple power law structure consistent with Lorentz and
conformal symmetry. The general forms have been used extensively before,
\cite{15,18}, and in momentum space we recall that they are, in the
Feynman gauge,
\begin{equation}
q(k) ~ \sim ~ \frac{\tilde{A}\kslash}{(k^2)^{\mu-\alpha}} ~~,~~
A_{\nu \sigma}(k) ~ \sim ~ \frac{\tilde{B}\eta_{\nu\sigma}}{(k^2)^{\mu-\beta}}
\end{equation}
where $\tilde{A}$ and $\tilde{B}$ are momentum independent amplitudes,
$\mu$ $=$ $\half d$ and the critical exponents $\alpha$ and $\beta$ are
defined as
\begin{equation}
\alpha ~=~ \mu - 1 + \half \eta ~~~,~~~ \beta ~=~ 1 - \eta - \chi
\end{equation}
Here the canonical dimension of each field is deduced from a dimensional
analysis of the kinetic terms and interactions of (1) and the anomalous terms
$\eta$ and $\chi$ are respectively the quark wave function renormalization
exponent and the $qqg$ vertex anomalous dimension. (At leading order in $1/\Nf$
there are no graphs involving ghosts, \cite{4}.) The presence of the non-zero
anomalous dimensions means that when considering subsequent orders in $1/\Nf$
only those graphs where the quark and gluon propagators are not dressed are
considered. From \cite{17,18} we recall that in the Feynman gauge
\begin{equation}
\eta_1 ~=~ \frac{2(\mu-1)^2 C_2(R) \eta^{\mbox{o}}_1}{(2\mu-1)(\mu-2)T(R)}
\end{equation}
where $\eta$ $=$ $\sum_{i=1}^\infty\eta_i/\Nf^i$ and $\eta^{\mbox{o}}_1$ $=$
$(2\mu-1)(\mu-2)\Gamma(2\mu)/[4\Gamma^2(\mu)\Gamma(\mu+1)\Gamma(2-\mu)]$.
Also, the combination $z$ $=$ $\tilde{A}^2\tilde{B}$ is required and
\cite{17,18}
\begin{equation}
z_1 ~=~ \frac{\mu\Gamma(\mu) \eta^{\mbox{o}}_1}{2(\mu-2)(2\mu-1)T(R)}
\end{equation}

We now recall the method to compute the anomalous dimensions of composite
operators with the critical propagators (6). First, one substitutes the lines
of fig. 1 with (6) and calculates the integral. However, one would discover
that the integral is divergent and thus requires regularization. This is
achieved by shifting the gluon exponent by an infinitesimal quantity $\delta$,
$\beta$ $\rightarrow$ $\beta$ $-$ $\delta$, \cite{13,15,18}. (In previous work,
we used the symbol $\Delta$ for the regularizing parameter but to avoid
confusion with $\Delta_\mu$ we use $\delta$ here.) When one computes with
non-zero $\delta$ now, each graph of fig. 1 will have the following formal
structure at $O(1/\Nf)$ after expanding in powers of $\delta$, \cite{24,17},
\begin{equation}
\frac{P}{\delta} ~+~ Q ~+~ R \ln p^2 ~+~ O(\delta)
\end{equation}
where $P$, $Q$ and $R$ depend on $\mu$ and $N_c$, and $p$ is the external
momentum flowing through the quark fields. The simple pole is absorbed
minimally by a conventional (critical point) renormalization, \cite{24}. This
leaves a term involving $R$ which would violate scaling symmetry in the limit
as one approaches criticality, $p^2$ $\rightarrow$ $\infty$. To avoid this
one notes that by general arguments the $\delta$-finite Green's function must
resum to the structure $(p^2)^{\gamma^{(n)}_{\cal O}(g_c)/2}$, \cite{24},
where $\gamma^{(n)}_{\cal O}(g_c)$ is related to the exponent we are interested
in, $\eta^{(n)}_1$, through, \cite{17},
\begin{equation}
\eta^{(n)} ~=~ \eta ~+~ \gamma^{(n)}_{\cal O}(g_c)
\end{equation}
and is what is obtained from the insertion of (5) in the graphs of fig. 1.
Therefore, by isolating and then summing the contributions from the $\ln p^2$
terms of each graph of fig. 1, we can deduce $\eta^{(n)}_1$ via (8) and (11).

Using (6) and the Feynman rules of \cite{4}, we find that the first graph of
fig. 1  contributes
\begin{equation}
- \, \frac{2\mu(\mu-1)^3C_2(R)\eta^{\mbox{o}}_1}{(\mu-2)(2\mu-1)(\mu+n-1)
(\mu+n-2)T(R)}
\end{equation}
to $\gamma^{(n)}_{\cal O}(g_c)$, whilst the second (together with its mirror
image) gives
\begin{equation}
\frac{4\mu(\mu-1)C_2(R)\eta^{\mbox{o}}_1}{(\mu-2)(2\mu-1)T(R)}
\sum_{l=2}^n \frac{1}{(\mu+l-2)}
\end{equation}
Therefore, from (8) and (11) we have
\begin{eqnarray}
\eta^{(n)}_1 &=& \frac{2C_2(R)(\mu-1)^2\eta^{\mbox{o}}_1}{(2\mu-1)(\mu-2)T(R)}
\left[ \frac{(n-1)(2\mu+n-2)}{(\mu+n-1)(\mu+n-2)} \right. \nonumber \\
&&+~ \left. \frac{2\mu}{(\mu-1)}\sum_{l=1}^n\frac{1}{(\mu+l-2)}
- \frac{2\mu}{(\mu-1)^2} \right]
\end{eqnarray}
which is the main result of this letter\footnote{Although the factor of
$(\mu-2)$ in the denominator of (14) suggests that $\eta^{(n)}_1$ behaves as
$1/\epsilon$ as $\epsilon$ $\rightarrow$ $0$ when $\mu$ $=$ $2$ $-$ $\epsilon$,
in fact $\eta^{\mbox{o}}_1$ $=$ $O(\epsilon^2)$ which means that $\eta^{(n)}_1$
and $\gamma^{(n)}_{\mbox{\footnotesize{NS}}}(g_c)$ are both $O(\epsilon)$.}.

There are several checks we can make on (14). First, in the case $n$ $=$ $1$,
the anomalous dimension $\gamma^{(1)}_{\mbox{\footnotesize{NS}}}(g)$ is zero,
\cite{4}, since then ${\cal O}_{\mbox{\footnotesize{NS}}}$ corresponds to a
conserved current. It is easy to verify from (14) that $\eta^{(1)}_1$ $=$ $0$
in agreement with this general result. Second, we can expand $\eta^{(n)}_1$ in
powers of $\epsilon$ in $d$ $=$ $4$ $-$ $2\epsilon$ dimensions and compare with
the $2$-loop result of [4-7] ie
\begin{eqnarray}
\gamma^{(n),\pm}_{\mbox{\footnotesize{NS}}}(g) &=& C_2(R)g \left[ 2S_1(n)
- \frac{3}{2} - \frac{1}{n(n+1)} \right] \nonumber \\
&&+~ \left[ \frac{C_2(R)T(R)\Nf}{9} \left( \frac{}{} \!\!\! 6S_2(n)
- 10S_1(n) \right. \right. \nonumber \\
&&+~ \left. \left. \frac{3}{4} + \frac{(11n^2+5n-3)}{n^2(n+1)^2}\right)
+ b^{\pm}(n) \right] g^2
\end{eqnarray}
where $S_l(n)$ $=$ $\sum_{i=1}^n 1/i^l$ and the coefficients $b^{\pm}(n)$ have
been given in [5-7] but are $O(1/\Nf^2)$ and their explicit form is not needed
for the present point. Substituting the critical coupling $g_c$ in (15) from
(4) and comparing with the $\epsilon$ expansion of (14) it is easy to verify
agreement with the two $O(1/\Nf)$ coefficients of (15). This is a highly
non-trivial check and establishes the correctness of (14) since we considered
only the leading order $1/\Nf$ graphs which are one loop. In other words the
conformal ans\"{a}tze and formalism correctly reproduce the large $\Nf$ bubble
sum contributions at higher order. More importantly having verified its
correctness we can now deduce subsequent terms in the perturbation series
albeit at leading order in $1/\Nf$. For example, if we set
\begin{equation}
\gamma^{(n)}_{\mbox{\footnotesize{NS}}}(g) ~=~ a_1 C_2(R)g
+ \sum_{i=2}^\infty a_i C_2(R) [T(R)\Nf]^{i-1} g^i
\end{equation}
for the leading order $1/\Nf$ part of the function, where $a_1$ and $a_2$ can
be read off from (15), then (14) implies that
\begin{eqnarray}
a_3 &=& \frac{2}{9}S_3(n) - \frac{10}{27}S_2(n) - \frac{2}{27}S_1(n)
+ \frac{17}{72} \nonumber \\
&&-~ \frac{[12n^4+2n^3-12n^2-2n+3]}{27n^3(n+1)^3} \\
a_4 &=& \frac{2}{27}S_4(n) - \frac{10}{81}S_3(n) - \frac{2}{81}S_2(n)
- \frac{2}{81}S_1(n) \nonumber \\
&&+~ \left[ \frac{4}{27}S_1(n) - \frac{2}{27n(n+1)} - \frac{1}{9}\right]
\zeta(3) + \frac{131}{1296} \nonumber \\
&&+~ \frac{[4n^6-12n^5-15n^4+10n^3+14n^2-n-3]}{81n^4(n+1)^4} \\
a_5 &=& \frac{2}{243}[3S_5(n) - 5S_4(n) - S_3(n) - S_2(n) - S_1(n)]
\nonumber \\
&&+~ \frac{\zeta(4)}{54}\left[ 4S_1(n) - 3 - \frac{2}{n(n+1)}\right]
+ \frac{323}{7776} \nonumber \\
&&+~ \frac{\zeta(3)}{486} \left[ 24S_2(n) - 40S_1(n) + 3
+ \frac{4[11n^2+5n-3]}{n^2(n+1)^2}\right] \nonumber \\
&&+~ \frac{[8n^7-8n^6-28n^5-5n^4+24n^3+13n^2-4n-3]}{243n^5(n+1)^5}
\end{eqnarray}
are the subsequent new coefficients, where $\zeta(q)$ is the Riemann zeta
function. Of course, $a_3$ will be an important check for the explicit $3$-loop
$\overline{\mbox{MS}}$ calculation of
$\gamma^{(n)}_{\mbox{\footnotesize{NS}}}(g)$ which is currently being
calculated for various moments, \cite{25}. To aid comparison of the $O(1/\Nf)$
part of the full three loop result of \cite{25}, we have evaluated $a_3$ for
even $n$, $2$ $\leq$ $n$ $\leq$ $22$, and collected the results in table 1. In
\cite{25}, the coefficient of the $n$ $=$ $8$ case was given and it is very
satisfying to record that there is exact agreement with the corresponding entry
of our table, which reinforces our confidence in the correctness of the
exponent (14), (after allowing for the different convention in defining the
coupling constant in this paper and an overall factor of 2 between the
definition of the renormalization group functions). The remaining values of
table 1 will occur in the explicit evaluation of
$\gamma^{(n)}_{\mbox{\footnotesize{NS}}}(g)$ in perturbation theory for other
$n$.

Finally, with (14) the $x$-behaviour of the $O(1/\Nf)$ part of the non-singlet
splitting function, $P_{\mbox{\footnotesize{NS}}}(g,x)$, of the
Altarelli-Parisi equations can now be examined, where $x$ is the Bjorken
scaling variable. This will be useful in the extension of the earlier $2$-loop
work of \cite{11,26}. It is defined as the Mellin transform of the anomalous
dimension,
\begin{equation}
\int_0^1 dx \, x^{n-1} \, P_{\mbox{\footnotesize{NS}}} (g,x) ~=~
- \, \mbox{\small{$\frac{1}{4}$}} \gamma^{(n)}_{\mbox{\footnotesize{NS}}}(g)
\end{equation}
where we use the conventions of \cite{27}.

We conclude by noting that we have given an alternative way of computing
anomalous dimensions of physical operators in QCD, which agrees with low order
results, and also provided the means to deduce the higher order structure. It
ought now to be possible to develop this large $\Nf$ formalism in two
directions. First, the $O(1/\Nf^2)$ corrections can, in principle, be deduced.
However, one would first need to know $\eta_2$, the quark anomalous dimension,
which has yet to be computed. Secondly, it may be possible to determine the
singlet operator anomalous dimensions. Although these mix under renormalization
that is not a significant problem since matrices of exponents have been deduced
in other contexts within the critical point programme, \cite{24}.

\vspace{1cm}
\noindent
{\bf Acknowledgement.} The author thanks D.J. Broadhurst for a useful
conversation. J.A.M. Vermaseren, S.A. Larin and T. van Ritbergen are also
thanked for providing a copy of and discussion on \cite{25}. Fig. 1 was drawn
with FeynDiagram version 1.17.
\newpage
{\begin{table}[h]
\hspace{3cm}
{ \begin{tabular}{c|r}
$n$ & $a_3$ \\
\hline
& \\
$2$ & $- \, \frac{28}{243}$ \\
& \\
$4$ & $ - \, \frac{384277}{1944000}$ \\
& \\
$6$ & $ - \, \frac{80347571}{333396000}$ \\
& \\
$8$ & $ - \, \frac{38920977797}{144027072000}$ \\
& \\
$10$ & $ - \, \frac{27995901056887}{95850016416000}$ \\
& \\
$12$ & $ - \, \frac{65155853387858071}{210582486065952000}$ \\
& \\
$14$ & $ - \, \frac{68167166257767019}{210582486065952000}$ \\
& \\
$16$ & $ - \, \frac{5559466349834573157251}{16553468064672354816000}$ \\
& \\
$18$ & $ - \, \frac{19664013779117250232266617}{56770118727793840841472000}$ \\
& \\
$20$ & $ - \, \frac{6730392290450520870012467}
{18923372909264613613824000}$ \\
& \\
$22$ & $ - \, \frac{16759806821032136669044226177}
{46048135637404510767879321600}$ \\
\end{tabular} }
\vspace{0.2cm}
{ \caption{Values of $a_3$ for various $n$.} }
\end{table}}
\pagebreak

\newpage
\noindent
{\Large {\bf Figure Captions.}}
\begin{description}
\item[Fig. 1.] Leading order graphs for $\eta^{(n)}_1$.
\end{description}
\end{document}